\documentclass[sigconf]{aamas} 

\usepackage{balance} 
\usepackage{subcaption}

\usepackage{siunitx}

\usepackage{multirow}

\setcopyright{ifaamas}
\acmConference[AAMAS '21]{Proc.\@ of the 20th International Conference on Autonomous Agents and Multiagent Systems (AAMAS 2021)}{May 3--7, 2021}{Online}{U.~Endriss, A.~Now\'{e}, F.~Dignum, A.~Lomuscio (eds.)}
\copyrightyear{2021}
\acmYear{2021}
\acmDOI{}
\acmPrice{}
\acmISBN{}

\acmSubmissionID{351}

\title{MAPFAST: A Deep Algorithm Selector for Multi Agent Path Finding using Shortest Path Embeddings}

\author{Jingyao Ren}
\affiliation{
  \institution{University of Southern California}
  \city{Los Angeles, California}}
  \email{jingyaor@usc.edu}

\author{Vikraman Sathiyanarayanan}
\affiliation{
  \institution{University of Southern California}
  \city{Los Angeles, California}}
  \email{sathiyan@usc.edu}

\author{Eric Ewing}
\affiliation{
  \institution{University of Southern California}
  \city{Los Angeles, California}}
  \email{ericewin@usc.edu}

\author{Baskin Senbaslar}
\affiliation{
  \institution{University of Southern California}
  \city{Los Angeles, California}}
  \email{baskin.senbaslar@usc.edu}

\author{Nora Ayanian}
\affiliation{
  \institution{University of Southern California}
  \city{Los Angeles, California}}
  \email{ayanian@usc.edu}

\begin{abstract}
Solving the Multi-Agent Path Finding (MAPF) problem optimally is known to be NP-Hard for both make-span and total arrival time minimization. While many algorithms have been developed to solve MAPF problems, there is no dominating optimal MAPF algorithm that works well in all types of problems and no standard guidelines for when to use which algorithm. 
In this work, we develop the deep convolutional network MAPFAST (Multi-Agent Path Finding Algorithm SelecTor), which takes a MAPF problem instance and attempts to select the fastest algorithm to use from a portfolio of algorithms. 
We improve the performance of our model by including single-agent shortest paths in the instance embedding given to our model and by utilizing supplemental loss functions in addition to a classification loss.
We evaluate our model on a large and diverse dataset of MAPF instances, showing that it outperforms all individual algorithms in its portfolio as well as the state-of-the-art optimal MAPF algorithm selector. We also provide an analysis of algorithm behavior in our dataset to gain a deeper understanding of optimal MAPF algorithms' strengths and weaknesses to help other researchers leverage different heuristics in algorithm designs.
\end{abstract}

\keywords{Multi-Agent Path Finding; Algorithm Selection; Deep Learning}

\newcommand{\BibTeX}{\rm B\kern-.05em{\sc i\kern-.025em b}\kern-.08em\TeX}

\begin{document}


\pagestyle{fancy}
\fancyhead{}


\maketitle 

\section{Introduction}
Multi-Agent Path Finding (MAPF) is the problem of finding collision-free paths for a team of agents traveling from start locations to goal locations on a map. 
Given a map as an undirected graph, an optimal MAPF algorithm computes paths for agents with the minimum cost, such that no two agents occupy the same vertex or traverse the same edge at the same timestep.
MAPF is applicable to a wide variety of problems, including automated warehouses, self-driving vehicles, and game engines.
In this work we study MAPF on 2D-grids.
Solving the MAPF problem optimally is known to be NP-Hard for both make-span and total arrival time minimization~\cite{yu2013structure, banfi2017intractability}.
Many types of optimal MAPF algorithms and their variants have been proposed, including Conflict Based Search (CBS)~\cite{sharon2015conflict}, a method based on branch-and-cut-and-price (BCP)~\cite{lam2019branch}, and a boolean satisfiability based algorithm (SAT)~\cite{surynek2016efficient}.
However, there is no optimal MAPF algorithm that dominates the others; each algorithm may perform well where others do poorly, and vice versa. This stems from the inherent differences of the algorithms.
For example, as we show in Section~\ref{sec:data_analysis}, CBS performs well on instances where average path lengths are small, while it performs poorly on instances where agents need to traverse long paths since the complexity of CBS increases as the number of potential conflicts between agents increase.
On the other hand, BCP performs well on instances with long path lengths, but is outperformed by CBS when the path lengths are small.
Moreover, many real-world instances may result in both long and short paths, increasing the difficulty of hand-picking an efficient algorithm.
It is also often unclear whether MAPF algorithms have strengths or weaknesses for different map types, since they are usually tested on a relatively small number of maps. 
Finally, map features such as the density and arrangement of obstacles are not the only factors that affect the solving speed of MAPF algorithms; for example, the number and configuration of agents also influence the performance of different algorithms.
The difficulty of hand-picking the fastest optimal MAPF algorithm for a particular instance necessitates the development of an algorithm selector that automatically selects the fastest optimal MAPF algorithm.

It has been shown that if a MAPF instance can be decomposed into multiple disjoint sub-problems, solving each sub-problem independently and combining their results can be significantly faster than solving the original problem with a MAPF algorithm~\cite{vsvancara2019combining}. Each sub-problem may have different map and agent characteristics, which will affect the runtime of whatever MAPF algorithm is chosen for each sub-problem. If it is possible to select an efficient algorithm for each sub-problem, we can greatly speed up the runtime of our overall algorithm. 
While many optimal MAPF algorithms can solve instances with hundreds of agents within several minutes~\cite{lam2019branch,felner2018adding,li2019improved}, thousands of agents in complicated maps quickly make finding a solution intractable. Moreover, the total number of the decomposed sub-problems may be too large, making it difficult and inefficient to hand-pick the best algorithms.
We therefore seek to develop an algorithm selector that can select which algorithm to use given a MAPF instance.

In this paper, we propose MAPFAST (Multi-Agent Path Finding Algorithm SelecTor), a novel automatic optimal MAPF algorithm selector based on a convolutional neural network (CNN) with inception modules~\cite{szegedy2015going}. We show that MAPFAST outperforms all existing optimal MAPF algorithm selectors on a large dataset of diverse MAPF instances. 
We propose two methods for improving the quality of our model over previous MAPF algorithm selectors: 
(1) augmenting MAPF instance encodings with single-agent shortest paths,
and (2) using supplemental loss functions to train our model in addition to a classification loss. 
We empirically show that single-agent shortest paths, regardless of map topology, contain enough information to  train an algorithm selector that outperforms any individual algorithm in our portfolio.
Compared to other existing algorithm selectors~\cite{sigurdson2019automatic, kaduri2020algorithm}, we introduce a more selective and up-to-date algorithm portfolio that includes optimization, satisfiability, and search based algorithms (namely BCP, SAT, CBS and CBSH~\cite{li2019improved, felner2018adding}). 
We also provide insights into the certain instance characteristics that may lead to algorithms to perform well in certain environments using a dataset of more than $24{,}000$ MAPF instances.
These insights may help researchers leverage different heuristics in future algorithm designs.

\section{Related Work}

Algorithm selection is the problem of selecting the best algorithm from a portfolio of algorithms to solve a given instance of a problem~\cite{kotthoff2016algorithm, rice1976algorithm}. 
Algorithm selection can be formulated as a prediction problem, where the goal is to predict the best algorithm from a portfolio for an input instance~\cite{smith2009cross}.
Such techniques have been successfully applied to many computational problems, including propositional satisfiability (SAT) and the traveling salesman problem (TSP)~\cite{xu2008satzilla, kerschke2018leveraging, xu2012evaluating}.

Although algorithm selection has been applied to many optimization problems, MAPF algorithm selectors have not been well studied in the literature.
Sigurdson et al.~\cite{sigurdson2019automatic} first proposed a classification model based on a convolutional neural network (CNN) by representing the MAPF instance as an RGB image.
Their model is a version of AlexNet~\cite{krizhevsky2012imagenet}, which is modified and retrained from image classification to try and predict the fastest solver given an image input. Their model demonstrated that it was possible to predict fastest algorithms for MAPF instances, although they only achieved relatively limited performance.
Kaduri et al.~\cite{kaduri2020algorithm} proposed two different models: one  based on CNNs using VGGNet~\cite{szegedy2015going}, and the other based on a tree-based learning algorithm named XGBoost~\cite{chen2016xgboost}. 
Their work uses a MAPF algorithm portfolio that includes only optimal search-based algorithms.
Based on our performance analysis in Section~\ref{sec:perf_analysis}, the best search-based algorithm in our algorithm portfolio, namely CBSH, is the fastest algorithm for only $30\%$ of test cases. 
Thus, omitting  non-search based algorithms handicaps the performance of an algorithm selector. 
Their best model, \emph{XGBoost~Cl}, requires hand-crafted MAPF features (e.g., number of agents, obstacle density) as input to their algorithm selector. Although the authors provide analysis of the impact of their hand-crafted features on the performance of their model, 
the performance for their algorithm selector may still be impaired by their small feature set that may not include some important features of an instance.


\section{Algorithm Portfolio}
The algorithm portfolio is a set of pre-selected candidate algorithms. When run, an algorithm selector will select a single algorithm from the portfolio to run on an instance and report the results of that algorithm. 
Given that MAPF instances are complex and diverse, a good algorithm portfolio must be diverse. Ideally, an algorithm in the portfolio should have strengths that cover for the weaknesses of the other algorithms.
Many optimal MAPF algorithms are built on top of similar approaches with different heuristics (e.g., CBS and its variants). 
We seek to include optimal MAPF algorithms that are inherently different from each other to find the best algorithms for a variety of MAPF instances.

We select the following four algorithms for our portfolio:

\begin{itemize}
    \item \textbf{Search-based}: Conflict-Based-Search (CBS)~\cite{sharon2015conflict} and its state-of-the-art variant with improved heuristics, CBSH~\cite{felner2018adding, li2019improved}; 
    \item \textbf{Optimization-based}: Branch-and-Cut-and-Price (BCP)~\cite{lam2019branch}, a method based on the decomposition framework for mathematical optimization;
    \item \textbf{Satisfiability-based}: A reduction of the MAPF problems to propositional satisfiability problem (SAT) \cite{surynek2016efficient}.
\end{itemize}

We also tested other algorithms such as 
EPEA*~\cite{goldenberg2014enhanced} and 
ICTS~\cite{sharon2013increasing}, but removed them due to their limited performance compared to the algorithms in our portfolio.

\subsection{Performance Analysis} \label{sec:perf_analysis}
Next, we show the capabilities and different characteristics of the algorithms in our portfolio by presenting a performance analysis for each individual algorithm. 
We use three metrics to evaluate the portfolio algorithms:
\begin{itemize}
    \item \textit{Accuracy} is the proportion of instances in which an algorithm is the fastest in the portfolio. 
    \item \textit{Coverage} is the proportion of the instances that an algorithm successfully solves within the time limit (5 minutes). 
    \item \textit{Runtime} is the overall time taken by the algorithm, in minutes, to solve all instances. 
    A default value of $5$ minutes is added to the runtime when the algorithm doesn't solve the input instance within time limit.
\end{itemize}

\begin{table}[tbh]
\caption{Performance analysis for portfolio algorithms on the entire dataset}
\label{tab:performance_analysis}
\centering
\resizebox{.95\columnwidth}{!}{
\begin{tabular}{lcccccc}
\midrule
\multirow{2}{*}{\textbf{Algorithm}} & \multirow{2}{*}{\textbf{Accuracy}} & \multirow{2}{*}{\textbf{Coverage}} & \multirow{2}{*}{\textbf{Runtime}} & \multicolumn{3}{c}{\textbf{Solving Time} }\\ \cmidrule{5-7} & & & & \textbf{Mean} & \textbf{Median} & \textbf{StdDev}\\
\midrule
 CBS & 0.1908 & 0.40 & 77,091 & 3.088 & 5.000 & 2.344\\
 CBSH & 0.2953 & \textbf{0.91} & \textbf{21,380} & 0.856 & 0.133 & 1.530\\
 BCP & \textbf{0.5129} & 0.90 & 22,265 & 0.892 & 0.050 & 1.631\\
 SAT & 0.0010 & 0.38 & 85,024 & 3.405 & 5.000 & 2.163\\
\midrule
 Oracle & 1.0 & 1.0 & 8,867 & 0.355 & 0.033 & 0.771\\
\midrule
\end{tabular}}
\end{table}

\begin{figure}[tbh]
    \begin{subfigure}{.75\columnwidth}
    \centering
    \includegraphics[width=\textwidth]{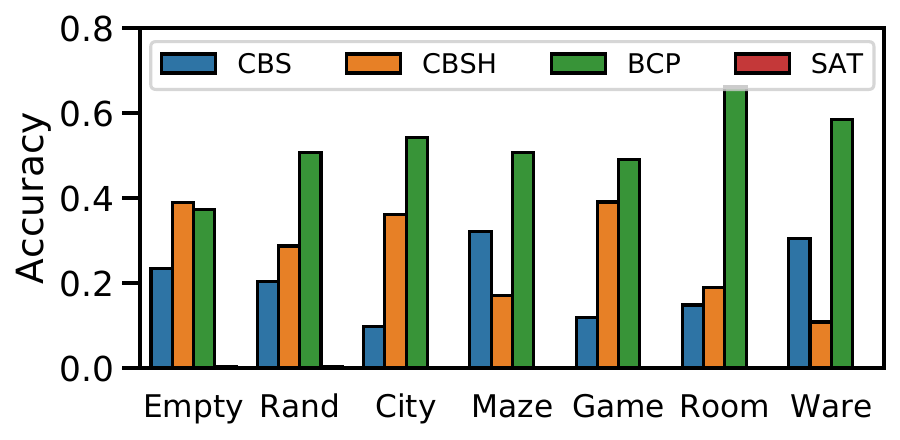} 
    \end{subfigure}
    \begin{subfigure}{.75\columnwidth}
    \centering
    \includegraphics[width=\textwidth]{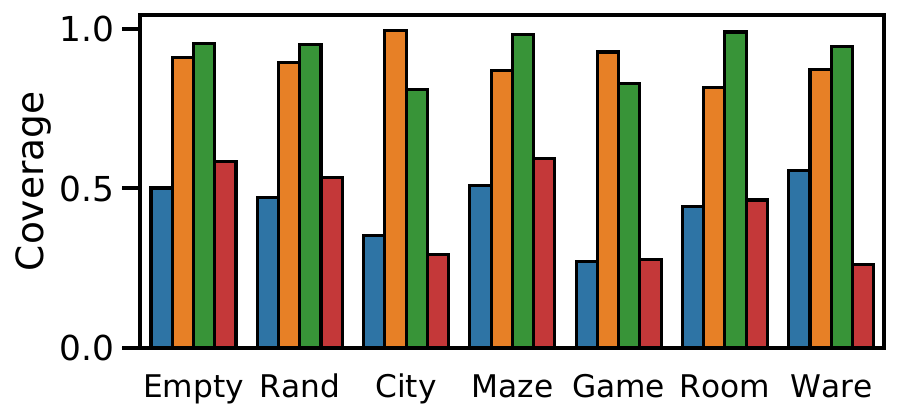} 
    \end{subfigure}
    \caption{Accuracy and coverage data for portfolio algorithms with respect to different map types. Random and Warehouse maps are labeled as Rand and Ware respectively.}
    \label{fig:ac_cov_map}
    \Description{Two bars plots showing the accuracy and coverage for portfolio MAPF algorithms on different map types.}
\end{figure}

\begin{table}[tbh]
    \caption{Accuracy and coverage data for game maps}
    \label{tab:simu_game}
    \centering
    \resizebox{.95\columnwidth}{!}{
    \begin{tabular}{c c c c c c c c c c}
    \midrule
    \multirow{2}{*}{\textbf{Map}} & \multicolumn{4}{c}{\textbf{Accuracy}} & & \multicolumn{4}{c}{\textbf{Coverage}}\\
    \cmidrule{2-5} \cmidrule{7-10}
    & CBS & CBSH & BCP & SAT & & CBS & CBSH & BCP & SAT\\
    \midrule
    brc202d & 0.1059 & \textbf{0.4677} & 0.4264 & 0 && 0.1719 & \textbf{0.9917} & 0.7318 & 0.1499\\
    orz900d & 0.1073 & \textbf{0.5181} & 0.3746 & 0 && 0.1677 & \textbf{0.9940} & 0.5755 & 0.0997\\
    den312d & 0.1378 & 0.2350 & \textbf{0.6272} & 0 && 0.3922 & 0.8379 & \textbf{0.9951} & 0.4668\\
    den520d & 0.1033 & \textbf{0.4813} & 0.4154 & 0 && 0.3091 & \textbf{0.9862} & 0.7608 & 0.2835\\
    lak303d & 0.1785 & 0.2714 & \textbf{0.5501} & 0 && 0.2459 & 0.8871 & \textbf{0.9508} & 0.3461\\
    ost003d & 0.1191 & \textbf{0.4839} & 0.3970 & 0 && 0.2208 & \textbf{0.9739} & 0.8102 & 0.2792\\
    \midrule
    game & 0.1185 & 0.3908 & \textbf{0.4906} & 0 && 0.2704 & \textbf{ 0.9271} & 0.8275 & 0.2757\\
    \midrule
    \end{tabular}}
\end{table}

\begin{figure}[tbh]
    \begin{subfigure}{.25\columnwidth}
    \centering
    \includegraphics[width=\textwidth]{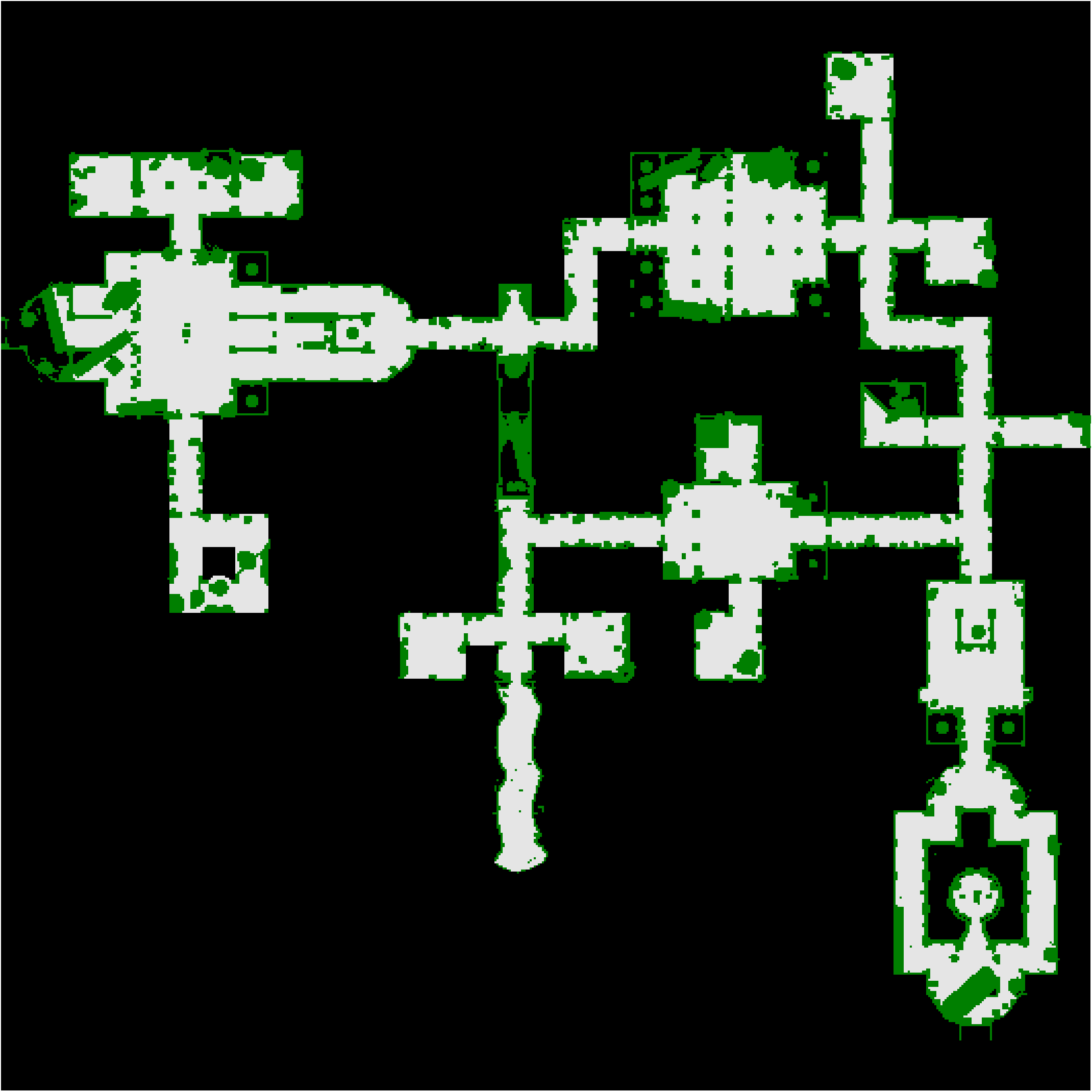}
    \end{subfigure}
    \begin{subfigure}{.25\columnwidth}
    \centering
    \includegraphics[width=\textwidth]{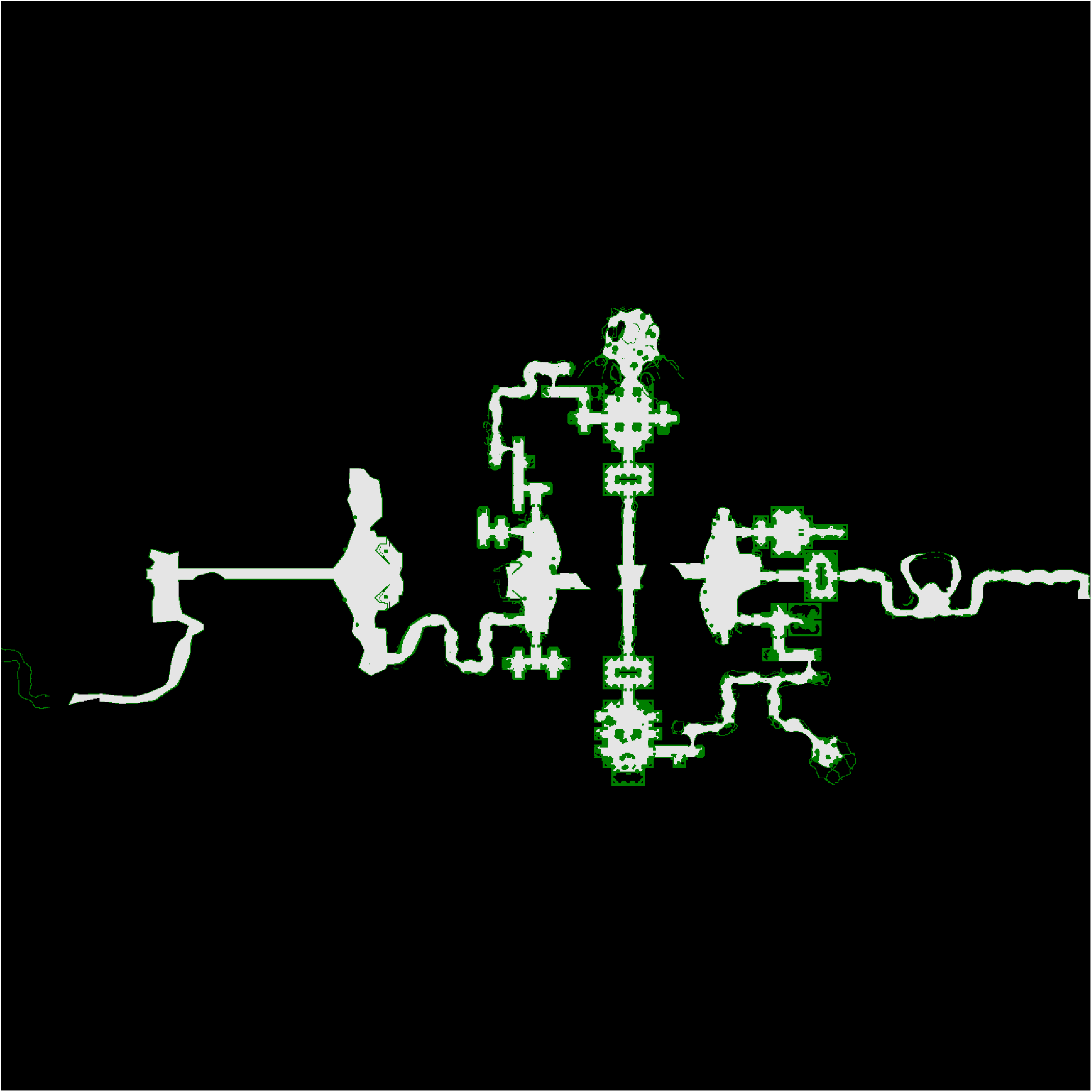}
    \end{subfigure}
    \begin{subfigure}{.25\columnwidth}
    \centering
    \includegraphics[width=\textwidth]{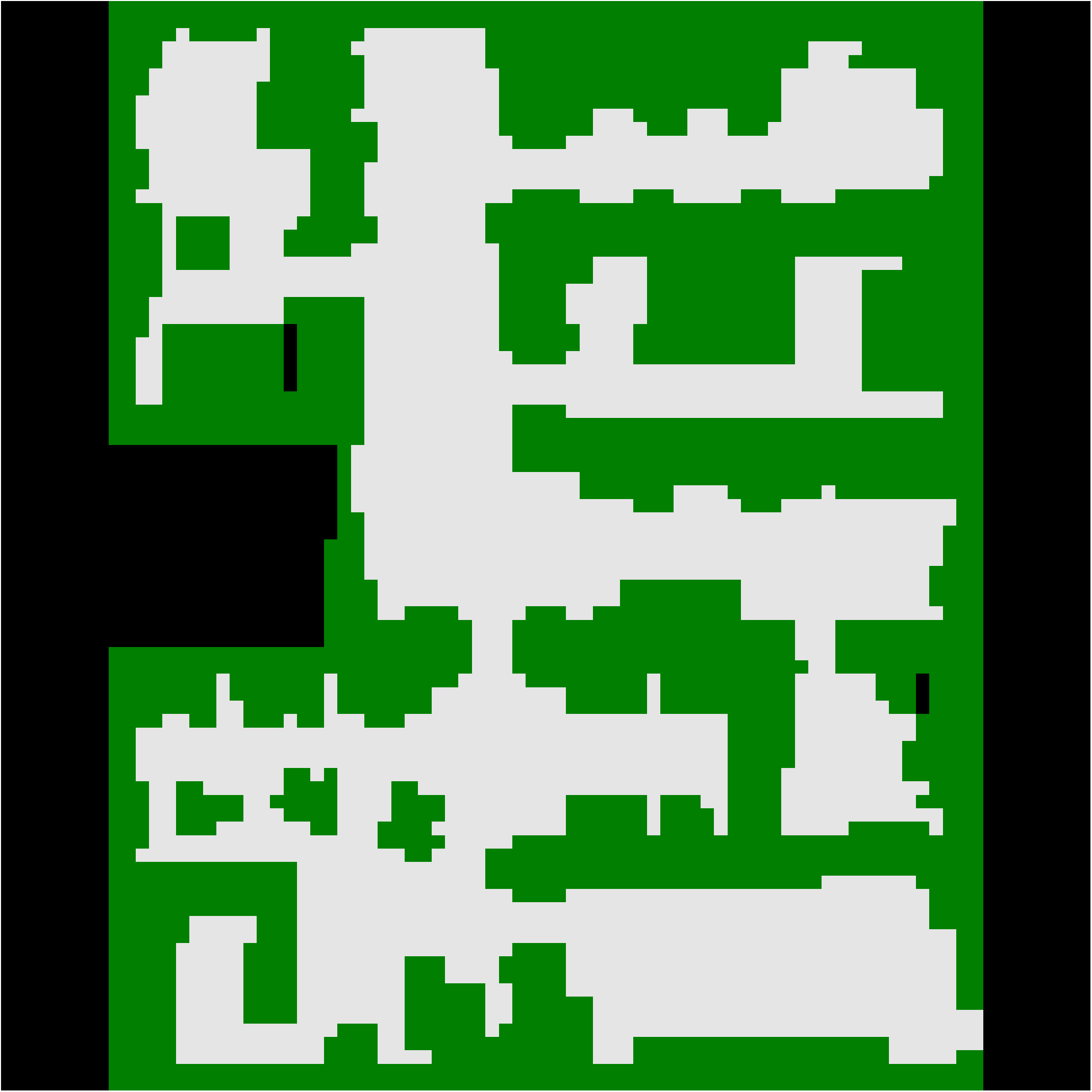}
    \end{subfigure}
    
    \begin{subfigure}{.25\columnwidth}
    \centering
    \includegraphics[width=\textwidth]{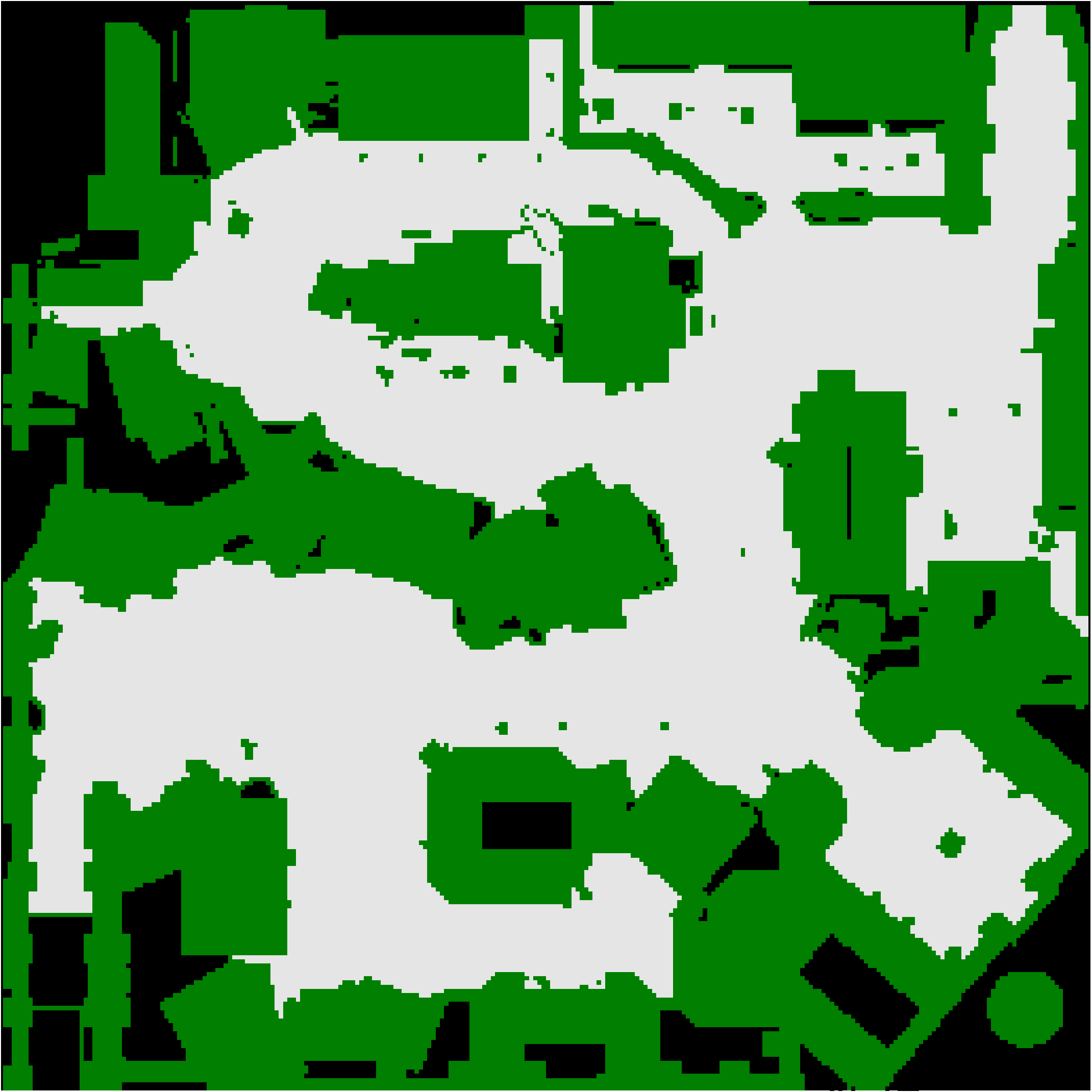}
    \end{subfigure}
    \begin{subfigure}{.25\columnwidth}
    \centering
    \includegraphics[width=\textwidth]{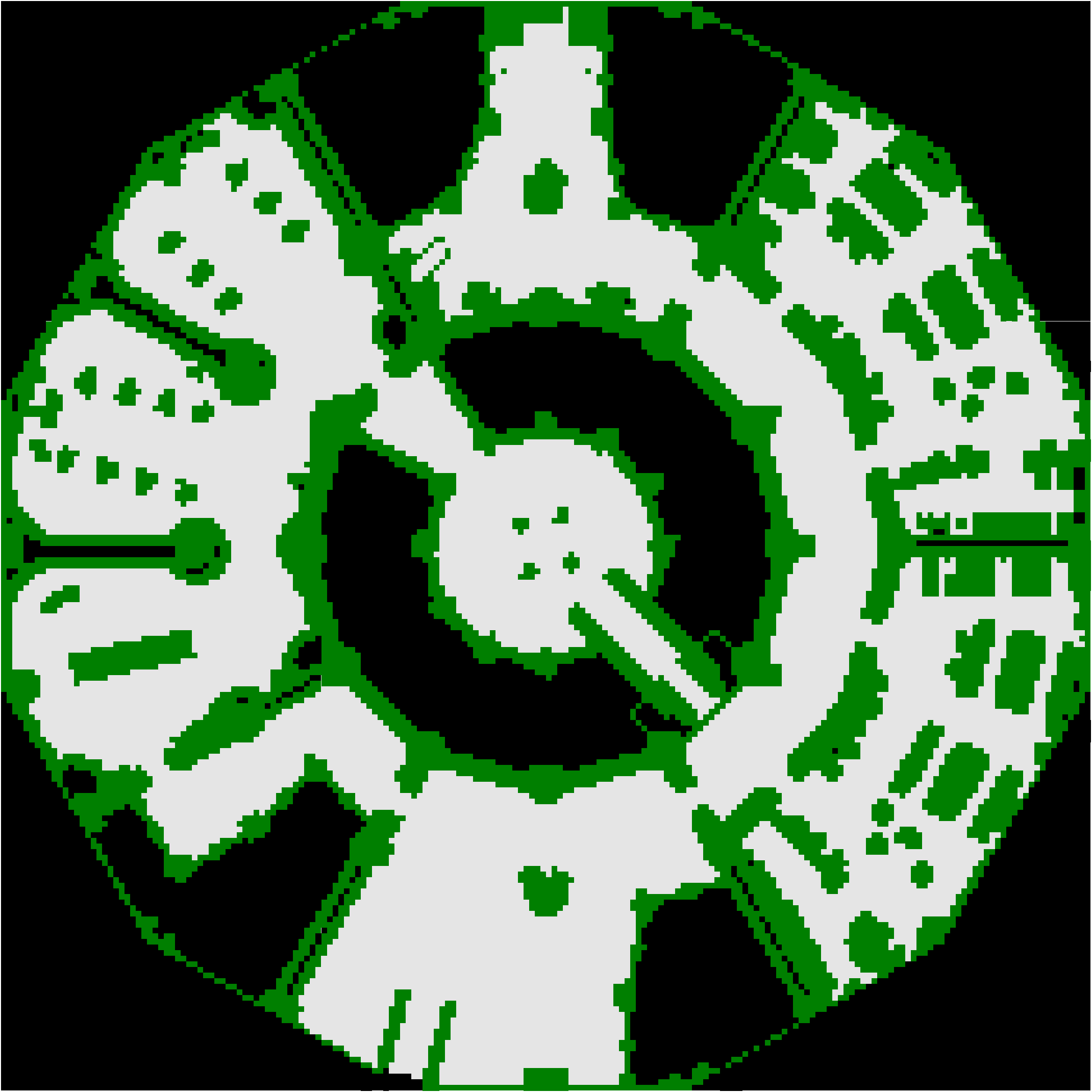}
    \end{subfigure}
    \begin{subfigure}{.25\columnwidth}
    \centering
    \includegraphics[width=\textwidth]{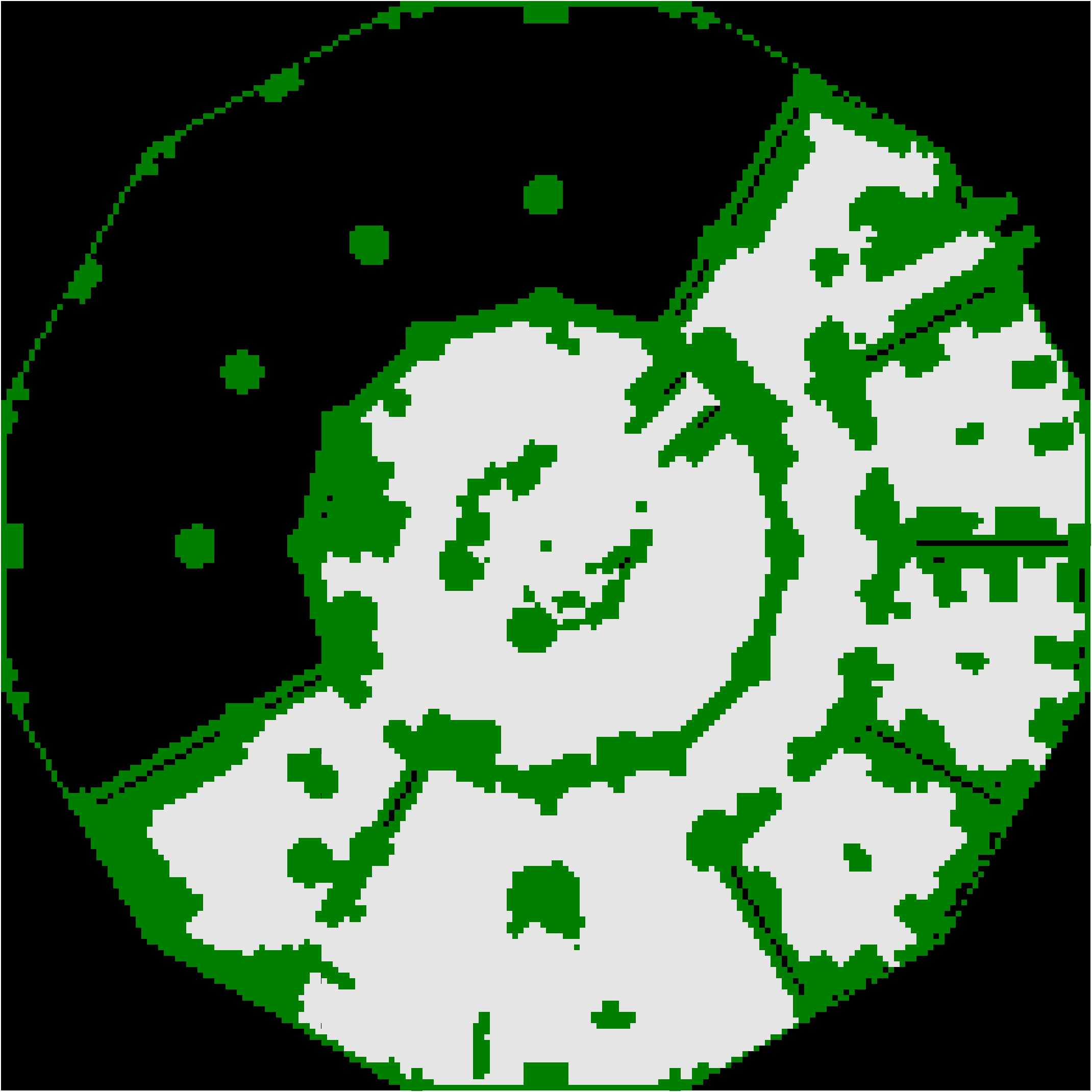}
    \end{subfigure}
    \caption{Maps in Table~\ref{tab:simu_game}. Left to right, first row: brc202d, orz900d, den312d; second row: den520d, lak303d, ost003d.}
    \label{fig:simu_game}
    \Description{Samples maps from MAPF Benchmark}
\end{figure}

We use the MAPF Benchmarks~\cite{stern2019mapf} to analyze our portfolio algorithms.  Our dataset of instances contains a wide variety of map types, including cities, video game maps, mazes, random maps and warehouses. 
When generating the instances, we only keep the instances that at least one algorithm from the portfolio can solve within the time limit ($5$ minutes).
We generate $24967$ solvable instances with varying number and distribution of agents. 
The results of the performance analysis are shown in Table~\ref{tab:performance_analysis}. 
We also include the mean, median, and standard deviation of the time needed (in minutes) for different algorithms to solve the instances.
BCP and CBSH are successful in solving $90\%$ of the instances. 
However, BCP, the algorithm that is fastest more often than any other algorithms in our portfolio, is only the fastest algorithm for $51\%$ of instances.
Selecting only BCP to solve all the instances would take more than twice as long as selecting the best performing algorithm for each instance (shown as Oracle in Table~\ref{tab:performance_analysis}). 
This further justifies the claim that there is no dominating optimal MAPF algorithm.

To answer whether a specific algorithm always performs well in certain types of maps, we present accuracy and coverage data with respect to the map types in Fig.~\ref{fig:ac_cov_map}. 
We also present the accuracy and coverage data for a subset of game map instances in Table~\ref{tab:simu_game}.  
Overall, BCP has the highest accuracy in the game maps.
However, neither BCP nor CBSH show dominant performance over each other for the instances on these individual maps.
CBSH outperforms BCP in brc202d, orz900d, den520d, ost003d but the gaps for accuracy are small. 
For den312d and lak303d, BCP is significantly better than CBSH in terms of accuracy. Even for the maps that have similar topology (e.g., den321d and den520d), the fastest algorithms can still be different.

Based on this data, we do not find a clear relationship between map types and algorithm performance. 
Some of the maps in Fig.~\ref{fig:simu_game} have both narrow corridors and open spaces,  making it challenging to manually choose the algorithm that works well on a certain map types (e.g., use CBS for maps with narrow corridors). 
Owing to the fact that map topologies 
are usually non-homogeneous, it is necessary to analyze MAPF instances on a case-by-case basis instead of categorizing them by map types.

\section{Instance Encoding}\label{sec:encoding}

\begin{figure}[t]
    \centering
    \begin{subfigure}{.3\columnwidth}
    \centering
       \includegraphics[width=\textwidth]{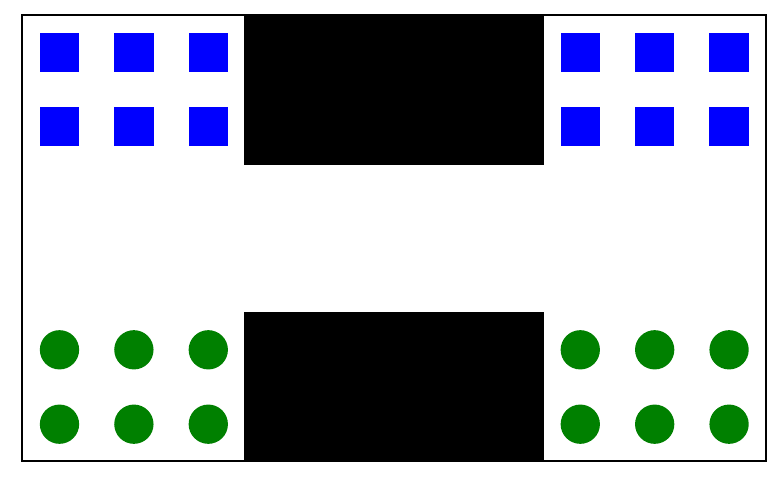}
    \caption{}
    \label{fig:narrow}
    \end{subfigure}
    \begin{subfigure}{.3\columnwidth}
    \centering
       \includegraphics[width=\textwidth]{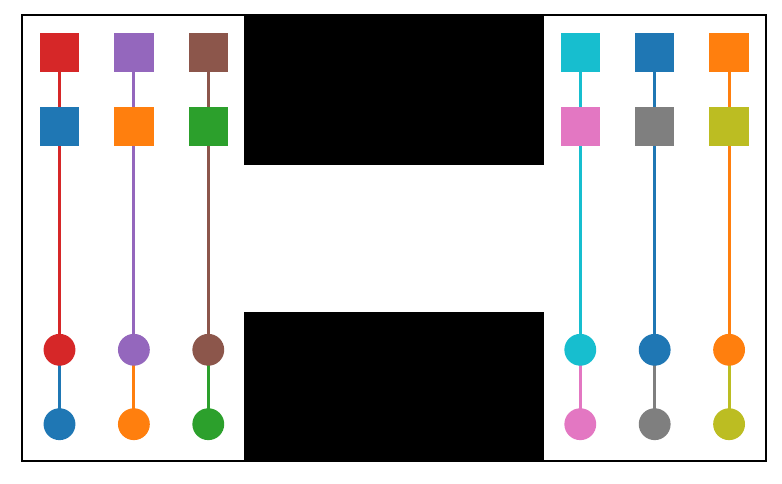}
    \caption{}
    \label{fig:narrow_easy}
    \end{subfigure}
    \begin{subfigure}{.3\columnwidth}
    \centering
       \includegraphics[width=\textwidth]{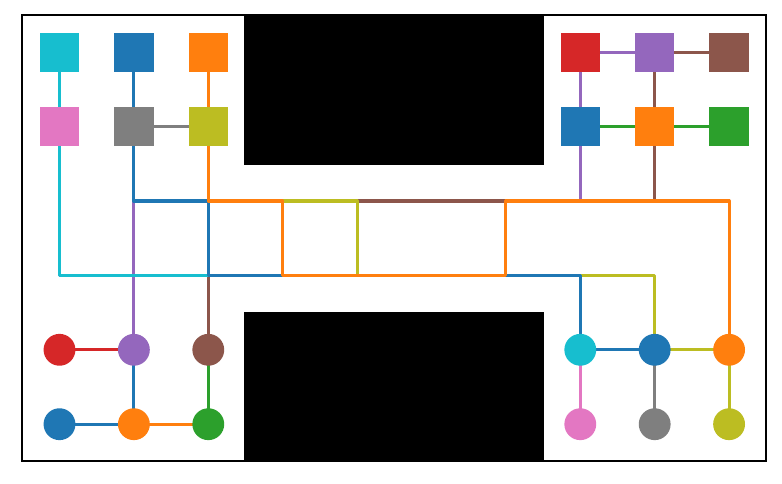}
    \caption{}
    \label{fig:narrow_hard}
    \end{subfigure}
    \caption{(a) MAPF instance marked only with start (green circles) and goal (blue squares) locations. (b)(c) Two different mappings of the start and goal locations with respect to the map in (a). Planned paths are marked in colored lines.}
    \Description{Figures showing that same distribution of start and goal locations may have different mappings of the start and goal locations such that will lead to very different solving difficulty.}
\end{figure}

\begin{figure}[t]
    \begin{subfigure}{.35\columnwidth}
    \centering
        \includegraphics[width=.8\textwidth]{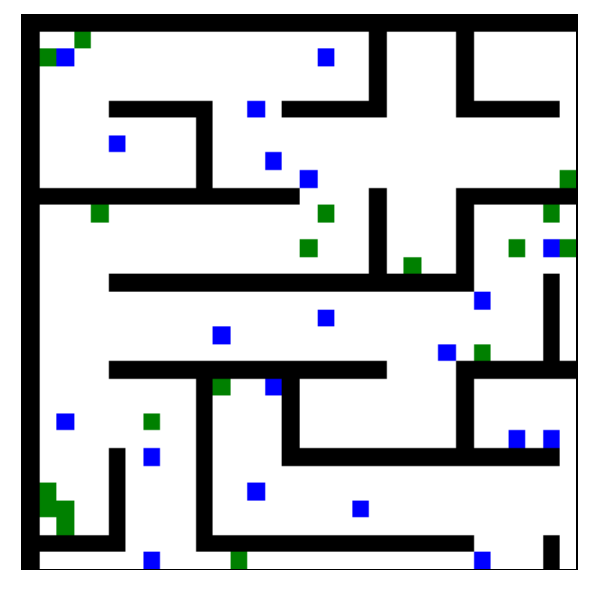}
    \caption{}
    \label{fig:ann1}
    \end{subfigure}
    \begin{subfigure}{.35\columnwidth}
    \centering
        \includegraphics[width=.8\textwidth]{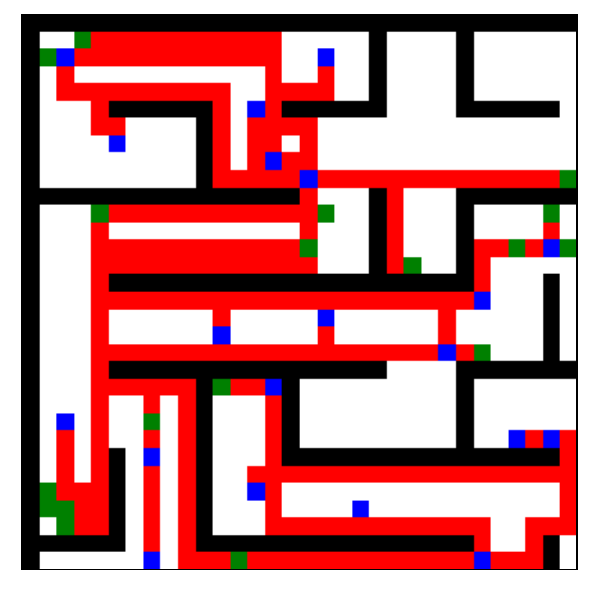}
    \caption{}
    \label{fig:ann2}
    \end{subfigure}
  \caption{Encoding an instance map with (a) start and goal locations and (b) single-agent shortest paths.}
  \Description{Two different ways of encoding MAPF instance, using just start and goal locations or additionally with single-agent shortest paths.}
\end{figure}

Existing MAPF algorithm selectors encode instances using either hand-crafted features~\cite{kaduri2020algorithm} or a 2D map with each cell marked as having a start or goal location or not~\cite{sigurdson2019automatic}. 
Encoding an instance using hand-crafted features cannot capture as much information as feeding the full map into a deep learning model. 
However, representing a variable number of agents and goals on a map is challenge. 
Sigurdson et al.~\cite{sigurdson2019automatic} encode agents and goals as binary features for each cell, meaning a node either has an agent/goal or does not. 
This method cannot distinguish between different permutations of agent start and goal positions: different assignments of start and goal locations to agents may lead to drastically different algorithmic performance. 
Take for instance the map shown in Fig.~\ref{fig:narrow} and two permutations of agents and goals in Fig.~\ref{fig:narrow_easy} and Fig.~\ref{fig:narrow_hard}.
CBS takes \SI{0.02}{s} and BCP takes \SI{0.03}{s} to solve the instance in Fig.~\ref{fig:narrow_easy}. 
With a different mapping for start and goal locations as shown in Fig.~\ref{fig:narrow_hard}, CBS doesn't finish within the five-minute time limit, but BCP successfully solves the instance in only \SI{0.33}{s}. 
Any model trained on binary encodings of agents and goals will not be able to differentiate between these two very different instances. 
We can improve the performance of our models by encoding into our inputs a mapping between agents and goals.
We will further justify this claim in the Section~\ref{sec:model_valid}.

We propose a new way of encoding MAPF instances that encodes a mapping between agents and their goals. 
In addition to marking the start and goal locations, we include another marking in our input which encodes \emph{single-agent shortest paths} from each agent to its goal.
A single-agent shortest path is an optimal path for an agent without considering collisions with other agents (Fig.~\ref{fig:ann2}). For every agent, we add only a single shortest path, despite the fact there may be many distinct shortest paths for every agent to its goal.
We encode the shortest paths on our map where each cell is marked if it lies on a shortest single-agent path.

We initially encoded our MAPF instance into a tensor with four layers. Each layer encoded a binary feature for each map cell: obstacle, agent, goal, shortest path. Models trained on this encoding performed relatively poorly in all metrics, perhaps requiring more training data than we generated. 
We then encoded our features in a different manner into a tensor with three layers by annotating every cell in our map with:
\begin{itemize}
    \item $[0, 0, 0]$ if the cell contains an obstacle
    \item $[1, 0, 0]$ if the cell lies on a shortest path from any agent to their goal
    \item $[0, 1, 0]$ if the cell is the starting location of an agent
    \item $[0, 0, 1]$ if the cell is the goal location of an agent
    \item $[0, 1, 1]$ if the cell is both a start and goal location
    \item $[1, 1, 1]$ if the cell is empty
\end{itemize}
Note that since every cell with a start/goal location is guaranteed to lie on a shortest path, we only mark that these cells contain a start/goal, we do not mark that they also lie on a shortest path since it is already implied by the presence of a start/goal. This is the encoding we use for our CNN-based models.


\section{Models}

We introduce the following algorithm selectors in this section: (1)~$\text{CNN}_\text{Class}$, which naively treats algorithm as a classification task, (2)~MAPFAST, an augmented version of $\text{CNN}_\text{Class}$ that achieves significantly better results, and (3)~G2V, a graph-embedding based model that offers insights into what information is required to perform algorithm selection on MAPF instances.

\subsection{$\text{CNN}_{\text{Class}}$}
In this work, we model algorithm selection as a classification task using a CNN. 
Our model takes as input an encoding of a MAPF instance and returns a prediction for the fastest algorithm in the portfolio. 
Inception modules are used to improve training speed and allow for a much deeper network than architectures like VGGNet~\cite{szegedy2015going}. 
Moreover, since the inception module contains multiple sizes of convolution kernels (Fig.~\ref{fig:cnn_long}), there is no need to decide the exact kernel size for each layer as the network learns which kernel to use.

The input to the model is a $320 \times 320 \times 3$ tensor of the encoding described in Section \ref{sec:encoding}
(smaller instances are padded with additional obstacles around the original map to make them $320 \times 320$).
The input is passed through three inception modules in our CNN (Fig.~\ref{fig:cnn_long}). 
Each of the inception modules is followed by a max-pooling layer (kernel size $3\times3$ with stride 3), a batch normalization layer and a rectified linear unit (ReLu) activation layer. 
Since the pre-trained inception network~\cite{szegedy2015going} uses the image size of $224 \times 224$ and our map size is $320 \times 320$, we cannot use the pre-trained weights, thus we train from scratch with the Adam optimizer~\cite{kingma2014adam}. 

After the three inception modules, the network outputs a feature vector of size $15488$. 
This is connected to a fully connected layer which outputs $200$ learned features.
This learned representation is then fed through a fully connected layer with a softmax activation function, which is the output of our model. The output is a prediction of which algorithm in the portfolio solves the input instance the fastest. We compute our classification loss $L_{Class}$ using categorical cross-entropy between our predictions and the ground truth fastest algorithm. This model is referred to as $\text{CNN}_{\text{Class}}$, as it is only trained with $L_{Class}$.

\subsection{MAPFAST}
To improve the quality of our predictions, we further augment our training process with two additional supplemental loss functions. We add a second output layer using four neurons with a sigmoid activation function to predict, for every algorithm, whether it will finish within the time limit or not. We compute our completion loss $L_{Comp}$ using cross-entropy loss between our second output layer and the ground truth algorithm completions.

\begin{figure*}[t]
\centering
\includegraphics[width=0.85\textwidth]{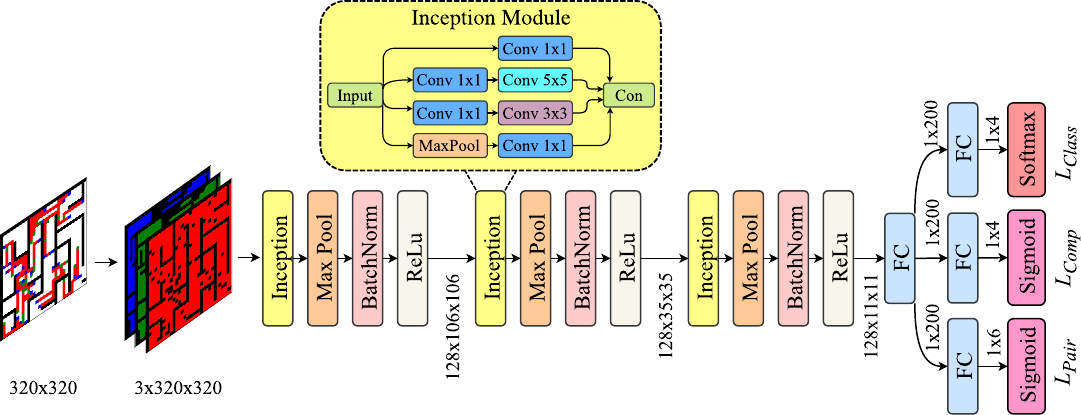}
\caption{The CNN architecture of MAPFAST.}
\Description{The CNN structure of our MAPF model has three inceptions modules.}
\label{fig:cnn_long}
\end{figure*}

We add an additional third output layer to predict pairwise relative performance of algorithms on an input instance. This is done using six output neurons. The first three output neurons predict whether BCP will be faster than CBS, CBSH, and SAT. The following two neurons predict whether CBS will be faster than CBSH and SAT. The final neuron predicts whether CBSH will be faster than SAT. Again, we compute our pairwise loss $L_{Pair}$ using cross-entropy. We train a model with the total loss $L_{Tot} = L_{Class} + L_{Comp} + L_{Pair}$ and refer to this combined model as MAPFAST (shown in Fig.~\ref{fig:cnn_long}).

\subsection{G2V}
For the previous models, we encode our MAPF instance as a tensor that contains information about the map as well as single-agent shortest paths. 
We now demonstrate that the single-agent paths alone, regardless of map topology, contain enough information to outperform any individual algorithm in our portfolio.
We utilize graph embedding techniques to convert the single-agent shortest paths for an instance into an embedding and train a model to make an algorithm prediction from this embedding.

For a MAPF instance, we construct our graph encoding as follows. 
First, we compute a single-agent shortest path for every agent. 
We then construct a graph consisting of only nodes which lie on these shortest paths. 
We add edges to all nodes which are adjacent in our MAPF instance. 
This (possibly disjoint) graph serves as the encoding of the MAPF instance. 
The size of our graph encoding is often significantly smaller than the size of the original instance map, as our graph encoding only contains nodes which lie along shortest paths.

After encoding our MAPF instance as a graph, we use the unsupervised graph embedding algorithm \textit{Graph2Vec}~\cite{narayanan2017graph2vec} to embed our graph into a vector. 
Graph2Vec takes as input a set of graphs and outputs a fixed-size vector representation for each graph. 
We feed Graph2Vec the graph representation of every instance in our dataset and it produces a vector of size $128$ for each instance. 
While Graph2Vec requires access to every graph in our dataset (including the test set) apriori, it does not have access to information on algorithm performances while generating embeddings and can be seen as a data preprocessing step. 
We train an XGBoost~\cite{chen2016xgboost} classifier on the embeddings generated from maps in our training set to predict the fastest algorithm in our portfolio.
This model is referred to as G2V in our results and is using only $L_{class}$, the cross-entropy loss to optimize the classification accuracy.

The drawback of this approach is that Graph2Vec needs all graphs \emph{before} any embedding is calculated. 
This means that, once trained, the model cannot be used to create a new graph embedding for an unseen instance. 
Therefore, it is not a deployable algorithm selector in any reasonable sense. However, we believe the results from this model are very informative. 
Our G2V model performs well despite only having access to nodes on shortest paths, potentially a very small fraction of the total number of nodes in the instance. 
In the shortest paths, there is no explicit information on map type, obstacle density, or map size, heuristics which have previously been used to select algorithms \cite{vsvancara2019combining, kaduri2020algorithm}.
Despite this lack of map information, our G2V model performs quite well, better than any single algorithm in our portfolio and close in performance to our CNN-based models, which have access to much more information.
This suggests that just information on single-agent shortest paths may be enough to distinguish the performances of our portfolio algorithms.


\section{Simulation Results} \label{sec:simu}

We evaluate the three models presented in the previous section: $\text{CNN}_\text{Class}$, MAPFAST, and G2V. 
Of the $24967$ instances generated from the MAPF Benchmarks~\cite{stern2019mapf} for our evaluation, we used 80\% for training, 10\% for validation and 10\% for testing. 
Before data collection, we built wrappers for the algorithms in our portfolio so that all the algorithms use a common method to read the input instances. 
The code can be found at this \href{https://github.com/USC-ACTLab/MAPFAST}{\textit{link}}.
We train our model for 5 epochs with a learning rate of $.001$, both determined by measuring performance on the validation set. All metrics reported in the following sections for algorithms and models are reported for the test set.
Training takes two hours using an RTX 2070 GPU.

\begin{table}[th]
\caption{Simulation results}
\label{t1}
\centering
\begin{tabular}{lccc}
\midrule
\textbf{Algorithm} & \textbf{Accuracy} &  \textbf{Coverage} & \textbf{Runtime}\\
\midrule
 CBS & 0.1888 & 0.41 & 7,714\\ 
 CBSH & 0.2810 & 0.90& \textbf{2,211}\\ 
 BCP & \textbf{0.5294} & \textbf{0.91} & 2,256\\ 
 SAT & 0.0008 & 0.38 & 8,548\\ 
\midrule
 XGBoost Cl & 0.6711  & 0.95 & 1,694\\
 $\text{CNN}_\text{Class}$ & 0.7118 & 0.95 & 1,560\\
 G2V & 0.7130 & 0.95 & 1,548\\
 MAPFAST & \textbf{0.7689} & \textbf{0.97} & \textbf{1,339}\\
\midrule
 Oracle & 1.0 & 1.0 &  917\\
\midrule
\end{tabular} 
\end{table}

We use the classification outputs of the models to select the fastest algorithm. To select an algorithm we take the argmax of the classification output, and select the corresponding algorithm (if the chosen algorithm fails to solve the instance, another algorithm is not selected).

As mentioned in Section~\ref{sec:perf_analysis}, we use accuracy, coverage, and runtime to evaluate the performance of portfolio algorithms. 
These metrics can also be used to analyze the performance of algorithm selectors, but with slightly different definitions. 

The \textit{Accuracy} metric gives the proportion of instances that the algorithm selector correctly selects the fastest algorithm. 
\textit{Coverage} is the proportion of instances where the algorithm selector selects an algorithm that solves the instance within the time limit. 
\textit{Runtime} is the overall time taken for the selected algorithms, in minutes, to solve all the problems in the test set.
A default value of 5 minutes was added to runtime when the algorithm didn't solve the input instance within time limit.

Table~\ref{t1} shows the results from evaluating our models on the $2484$ MAPF instances in the test set. 
In the first four rows, we report results for using a single algorithm on all input instances. Our experiments show that BCP and CBSH were successful in solving $90\%$ of the input instances. 
However, BCP, the best individual algorithm in accuracy and coverage, is  the fastest for only $53\%$ of instances and takes more than twice as long as selecting the fastest algorithm for each instance (shown as Oracle in Table~\ref{t1}).

The second part of the table shows the comparison of a state-of-the-art MAPF algorithm selector, XGBoost Cl~\cite{kaduri2020algorithm}, and our approach. To generate these results, we train XGBoost Cl with our algorithm portfolio and dataset. 
MAPFAST successfully selects the fastest algorithm for $77\%$ of the input instances and had coverage of $97\%$, outperforming XGBoost Cl, which had $67\%$ accuracy and $95\%$ coverage. 
Our Model G2V had a performance comparable to our $\text{CNN}_\text{Class}$ model, with $71\%$ accuracy and $96\%$ coverage, also outperforming XGBoost Cl.

The total runtime for the algorithms chosen by our models are significantly less than using the same algorithm for every instance. On average, it takes 1 second to annotate one input instance with single-agent shortest paths and 0.01 second for the trained model to select the fastest algorithm, which are negligible to the average runtime of the best portfolio algorithm (i.e., CBSH has an average runtime of $53$ seconds).
Our models also have a remarkable improvement of accuracy compared to all of the individual algorithms, further justifying our approach.

\begin{table}[th]
\caption{Actual and predicted coverage for MAPFAST}
    \centering
    \begin{tabular}{lcccc}
    \midrule
     & CBS & CBSH & BCP & SAT\\
    \midrule
    Actual Coverage & 0.41 & 0.90 & 0.91  & 0.38\\
    Predicted Coverage & 0.42 & 0.89 & 0.87 & 0.40\\
    Recall & 0.90 & 0.95 & 0.95 & 0.91\\
    Correctness & 0.91 & 0.93 & 0.92 & 0.91\\
    \midrule
    \end{tabular}
    \label{tab:coverage_analysis}
\end{table}

\newcommand{\mS}{\mathcal{S}}
\newcommand{\mQ}{\mathcal{Q}}
\newcommand{\mT}{\mathcal{T}}
\newcommand{\tmS}{\tilde{\mathcal{S}}}
\newcommand{\tmQ}{\tilde{\mathcal{Q}}}

We use the following method to further analyze the performance of the four output neurons in MAPFAST that use $L_{Comp}$ loss to predict if an algorithm solves a given input instance or not.
Let $\mT$ be the set of all test instances.
For a particular algorithm, let $\mS$ be the set of test instances that it can solve within the time limit, and $\mQ$ be the set of test instances it cannot solve within the time limit such that $\{\mS, \mQ\}$ is a partition of $\mT$.
Let $\tmS$ be the set of test instances our model predicts as solvable by the algorithm within the given time limit, and $\tmQ$ be the instances that our model predicts as not solvable by the algorithm within the given time limit.
$\{\tmS, \tmQ\}$ is another partition of $\mT$.
The first row of Table~\ref{tab:coverage_analysis} shows the actual coverage of the algorithms in the portfolio, i.e. $\frac{|\mS|}{|\mT|}$.
The second row shows the predicted coverage of our model for each algorithm, i.e. $\frac{|\tmS|}{|\mT|}$.
The third row lists the recall of our model, which is the fraction of solvable instances that our model predicts as solvable: $\frac{|\mS \cap \tmS|}{|\mS|}$.
The final row lists the correctness of our model, which is the fraction of correct outputs: $\frac{|(\mS \cap \tmS) \cup (\mQ \cap \tmQ)|}{|\mT|}$.
Our model predicts whether an algorithm solves an instance or not with at least $91\%$ correctness for each algorithm.
This suggests that the neural network learns the inherent behavior of algorithms for the given MAPF instances.

\subsection{Custom Scoring}
\begin{table*}[t]
\centering
\caption{Custom score results}
\label{tab:custom_score}
\begin{tabular}{lcccccccccc}
\midrule
\textbf{Algorithm} & SAT & CBS & CBSH & BCP & $\text{CNN}_{\text{Agents}}$ & XGBoost Cl & $\text{CNN}_\text{Class}$ & G2V  & MAPFAST & Oracle\\
\midrule
\textbf{Custom Score} & 28.23 & 78.93 & 222.26 & 280.41 & 287.50 & 288.12 & 296.11 & 299.99 & \textbf{320.43} & 382.03\\
\midrule
\end{tabular} 
\end{table*}

We also use an additional metric, the \emph{speed award}~\cite{van2005purse}, which provides more information about relative performance among different algorithms, to further analyze our models. 
This metric gives a greater reward for solving tasks that not every algorithm solves and a smaller reward to fast algorithms when every algorithm takes around the same amount of time. 
For different algorithm selectors, it gives greater rewards for the models that correctly choose the fastest algorithm when other models fail to do so. 
It therefore provides more information on the relative performance of algorithms and algorithm selectors than the accuracy and the cumulative runtime.

To calculate the speed award, we first compute the speed factor:
\begin{equation*}
    \mathit{speedFactor}(p, a_i) = \frac{300}{1 + \mathit{timeUsed}(p, a_i)},
\end{equation*}
where $\mathit{timeUsed}(p, a_i)$ is the time taken by the algorithm $a_i$ to solve instance $p$ and $300$ is the time limit for each instance. 
The speed factor shows how fast an algorithm can solve an instance. The faster an algorithm is, the higher the speed factor will be. 
If algorithm $a_i$ fails to solve the instance $p$, the speed factor is set to $0$.

Once we have the speed factor of all algorithms for a problem instance $p$, we compute the speed award for each algorithm $a_i$ to solve instance $p$ as follows:
\begin{equation*}
    \mathit{speedAward(p, a_i)} = \frac{\mathit{speedFactor}(p, a_i)}{\sum_{a_j \in \mathtt{algorithms}} \mathit{speedFactor}(p, a_j)}.
\end{equation*}
Here, $\mathtt{algorithms}=\{$BCP, CBS, CBSH, SAT, XGBoost Cl, G2V, $\text{CNN}_\text{Class}$, $\text{CNN}_\text{Agents}$, MAPFAST, Oracle$\}$. 
The speed award for an algorithm has a higher value if the algorithm solves the instance faster than other algorithms.

The final score for an algorithm on a set of problem instances is given by
\begin{equation}\label{eq:score}
    \mathit{score(a_i)} = \sum_{\forall p} \mathit{speedAward}(p, a_i).
\end{equation}

The custom score metric provides more information about the relative performance between different algorithms selectors than just using runtime metric. In particular, if the algorithms selected by different algorithm selectors have very similar runtime, then similar scores will be granted to these algorithm selectors instead of giving all the credits to the fastest algorithm.

The calculated custom scores are shown in Table \ref{tab:custom_score}. 
Oracle is the model that always selects the fastest algorithm. $\text{CNN}_\text{Agents}$ is used for model validation which will be introduced in Section~\ref{sec:model_valid}.

The best model should have the highest custom score.
Based on the results in Table~\ref{tab:custom_score}, all of the algorithm selectors outperform the portfolio algorithms. 
Moreover, all of our models outperform the state-of-the-art model, XGBoost Cl. MAPFAST is ranked as the best algorithm selector by \textit{speedAward}.

\subsection{Model Validation} \label{sec:model_valid}

\begin{table}[t]
\centering
\caption{Ablation study for CNN models}
\label{tab:ablation}
\begin{tabular}{lccc}
\midrule
\textbf{Algorithm} & \textbf{Accuracy} &  \textbf{Coverage} & \textbf{Runtime}\\
\midrule
 $\text{CNN}_\text{Agents}$ & 0.6710  & 0.93 & 1,796\\
 $\text{CNN}_\text{Class, Pair}$ & 0.7061 & 0.96 & 1,605\\
 $\text{CNN}_\text{Pair, Comp}$ & 0.7061& 0.96& 1,483\\
 $\text{CNN}_\text{Class}$ & 0.7118 & 0.95 & 1,560\\
 $\text{CNN}_\text{Pair}$ & 0.7154 & 0.94 & 1,757\\
 $\text{CNN}_\text{Class, Comp}$ & 0.7335 & 0.96 & 1,471\\
 MAPFAST & \textbf{0.7689} & \textbf{0.97} & \textbf{1,339}\\
\midrule
 Oracle & 1.0 & 1.0 &  917\\
\midrule
\end{tabular} 
\end{table}

We performed an ablation study to analyze architectural and design choices in our network by training variants of our model with different combinations of our loss functions.
We denote the loss functions we used as subscripts to our model, for example $\text{CNN}_\text{Class, Comp}$ is our model trained with only the classification loss $L_{Class}$ and the completion loss $L_{Comp}$. Note that with this notation, MAPFAST is equivalent to $\text{CNN}_{\text{Class, Pair, Comp}}$, but we refer to it as MAPFAST for simplicity.
For models that have classification outputs, we select the algorithm with the highest value in the classifier output.
For models that have pairwise comparison outputs and are trained with the pairwise performance loss $L_{Pair}$, but don't have classifier outputs, we select an algorithm according to the predicted relative performance of each algorithm.
We do not evaluate a model that only uses $L_{Comp}$ as there is not a reasonable way to select an algorithm from predictions of algorithm completion.
Additionally, we trained a model that only used agents' start and goal locations and did not include single-agent shortest paths, referred to as $\text{CNN}_\text{Agents}$.
The $\text{CNN}_\text{Agents}$ takes the same input encoding as \cite{sigurdson2019automatic}, however due to our different map size, we trained a new model from scratch. 
Additionally, we saw a performance increase by using inception modules rather than their model architecture, so we present results for  $\text{CNN}_\text{Agents}$ with the same model architecture as MAPFAST, but with different instance encoding. Our results for all models are presented in Table~\ref{tab:ablation}.

All models in Table~\ref{tab:ablation} outperform each individual algorithm in our portfolio (cf. Table \ref{t1}).  Training the model to predict algorithm completion has mixed impact on accuracy, but improves coverage and runtime scores. Interestingly, combining $L_{Class}$ and $L_{Pair}$ decreases accuracy, until $L_{Comp}$ is also added to make MAPFAST, which causes accuracy to increase $5\%$. Our model MAPFAST, using all three loss functions, has best performance in all three metrics.

We have two possible explanations for why the additional loss functions improved performance of MAPFAST:
(1) The classification loss function can be noisy, since sometimes the difference between solution times of the two fastest algorithms is quite small. The additional loss functions may smooth out this noise and prevent the model from reaching a local minima.
(2)  $L_{comp}$ helps the model avoid selecting algorithms that do not finish by predicting whether an algorithm finishes within the time limit or not.  $L_{pair}$ encodes more about the entire ordering, not just the fastest algorithm. Both can be seen as a means of extracting more information from each instance than using classification accuracy alone.

The custom score of $\text{CNN}_\text{Agents}$ in Table~\ref{tab:custom_score} further shows its limitation and the necessity of using shortest path embedding.
Albeit having the same model architecture, the score for MAPFAST ($320.43$) is remarkably higher than $\text{CNN}_\text{Agents}$ ($287.50$), which shows the significant enhancement that the shortest path embedding can bring.

\begin{figure*}[tbh]
    \centering
    \begin{subfigure}{.24\textwidth}
    \centering
    \includegraphics[width=\textwidth]{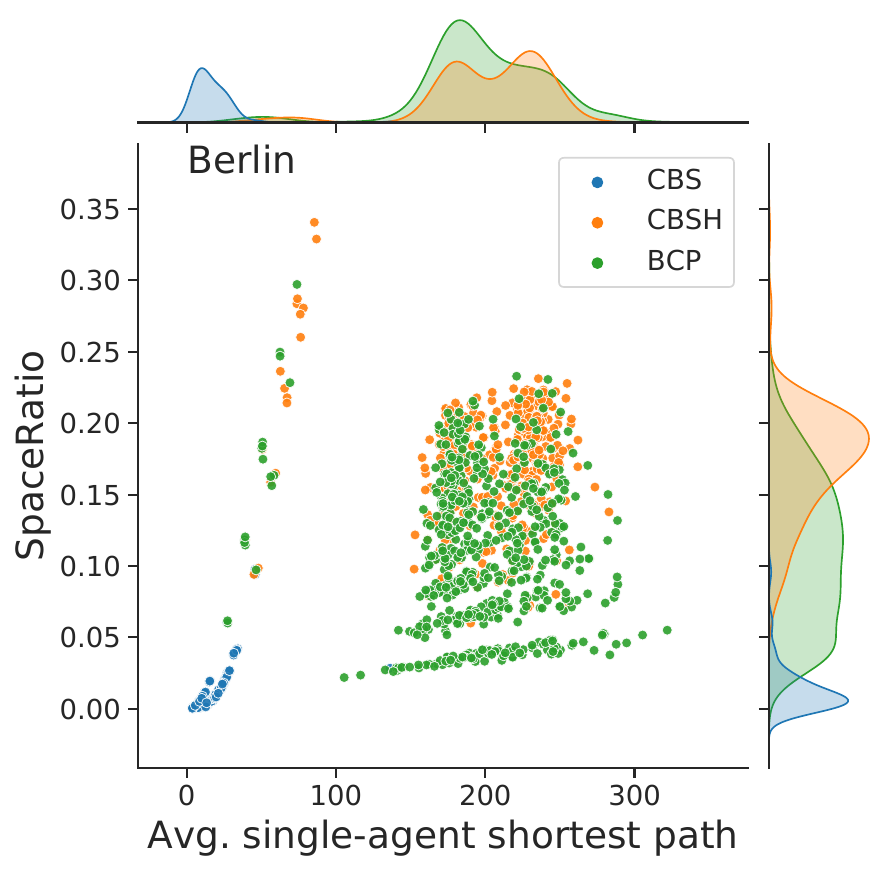}
    \caption{}
    \label{fig:apl_sp_berlin}
    \end{subfigure}
    \begin{subfigure}{.24\textwidth}
    \centering
    \includegraphics[width=\textwidth]{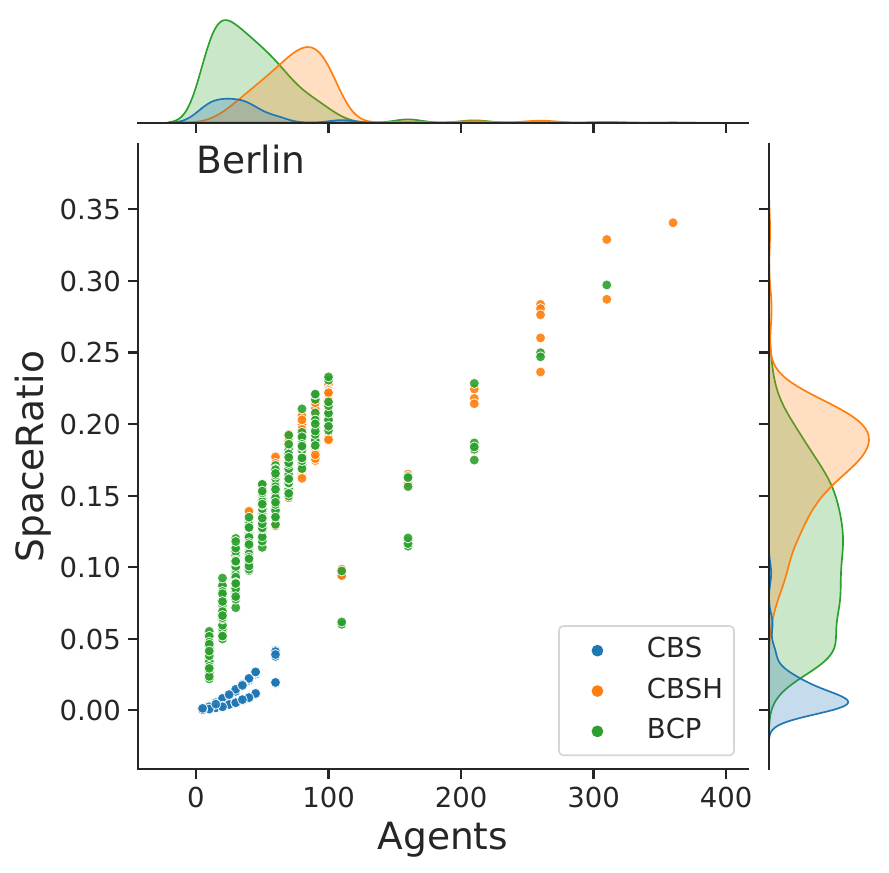}
    \caption{}
    \label{fig:ag_sp_berlin}
    \end{subfigure}
    \centering
    \begin{subfigure}{.24\textwidth}
    \centering
    \includegraphics[width=\textwidth]{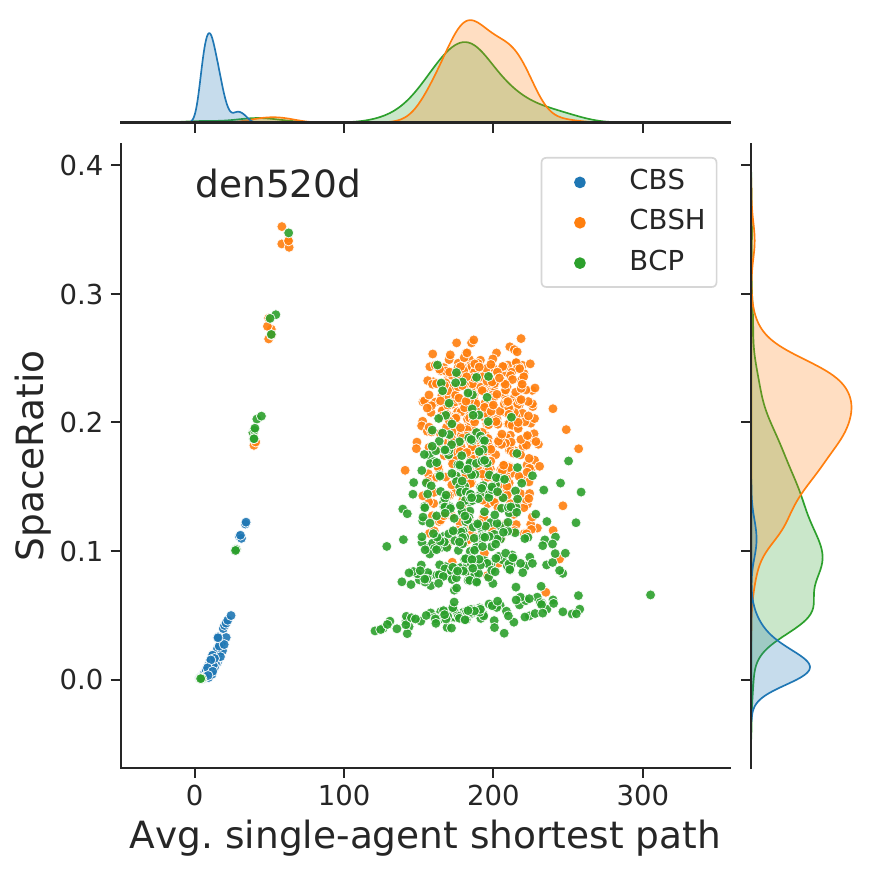}
    \caption{}
    \label{fig:apl_sp_den520d}
    \end{subfigure}
    \begin{subfigure}{.24\textwidth}
    \centering
    \includegraphics[width=\textwidth]{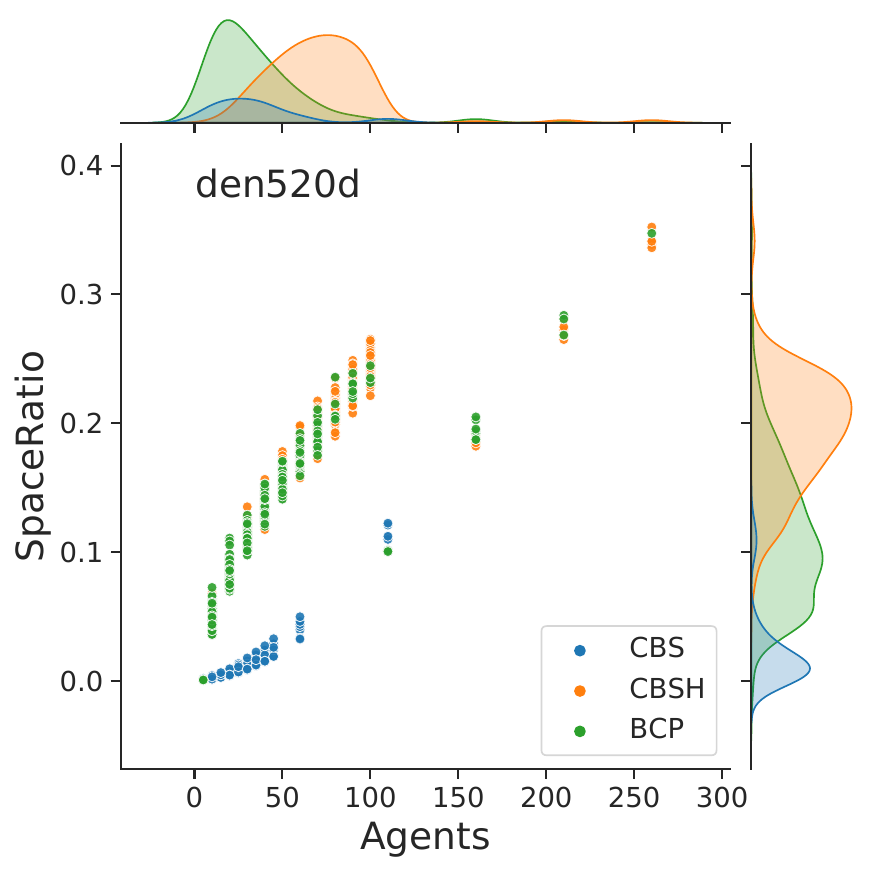}
    \caption{}
    \label{fig:ag_sp_den520d}
    \end{subfigure}
    
    \centering
    \begin{subfigure}{.15\textwidth}
    \centering
    \includegraphics[width=\columnwidth]{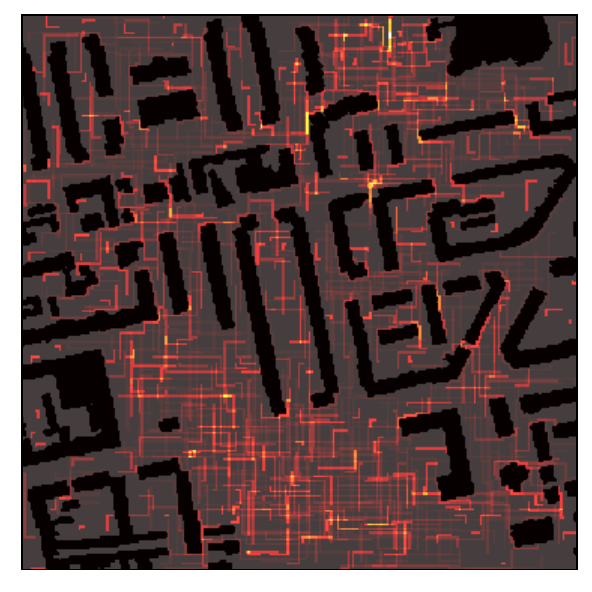}
    \caption{CBS}
    \label{fig:heatmap_1}
    \end{subfigure}
    \begin{subfigure}{.15\textwidth}
    \centering
    \includegraphics[width=\columnwidth]{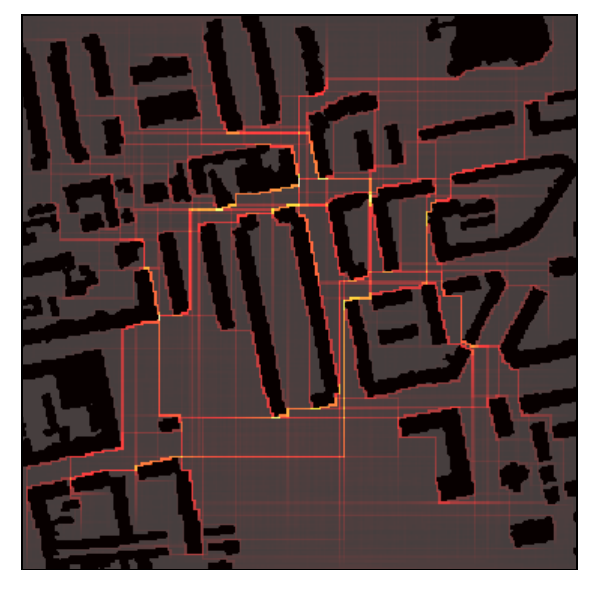}
    \caption{CBSH}
    \end{subfigure}
    \begin{subfigure}{.15\textwidth}
    \centering
    \includegraphics[width=\columnwidth]{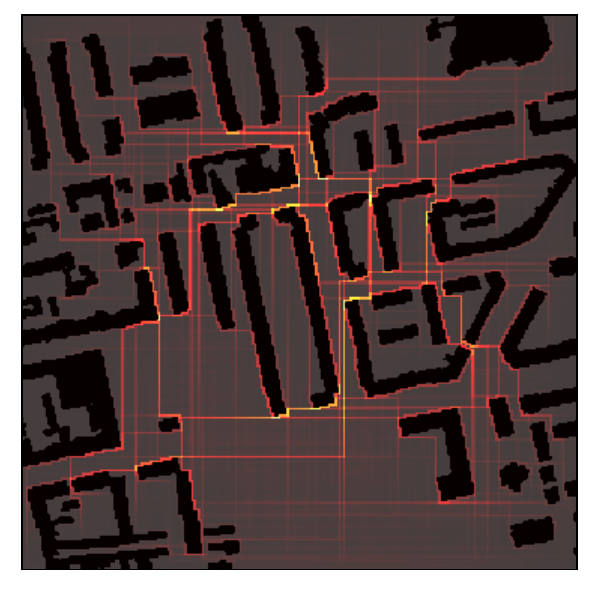}
    \caption{BCP}
    \end{subfigure}
    \begin{subfigure}{.15\textwidth}
    \centering
    \includegraphics[width=\columnwidth]{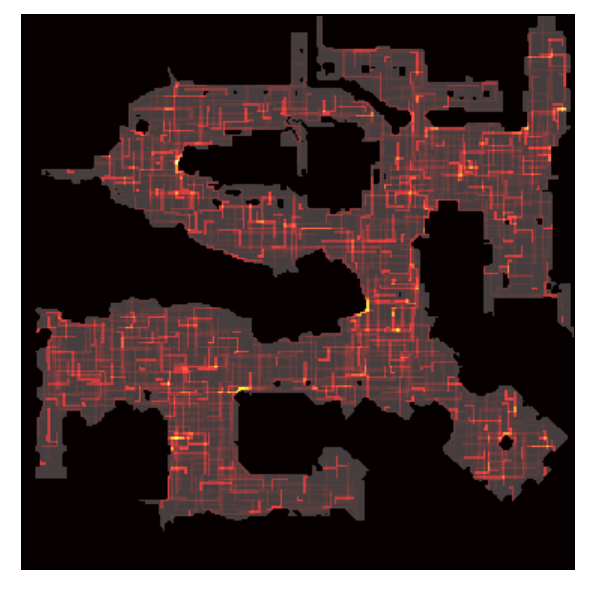}
    \caption{CBS}
    \label{fig:heatmap_4}
    \end{subfigure}
    \begin{subfigure}{.15\textwidth}
    \centering
    \includegraphics[width=\columnwidth]{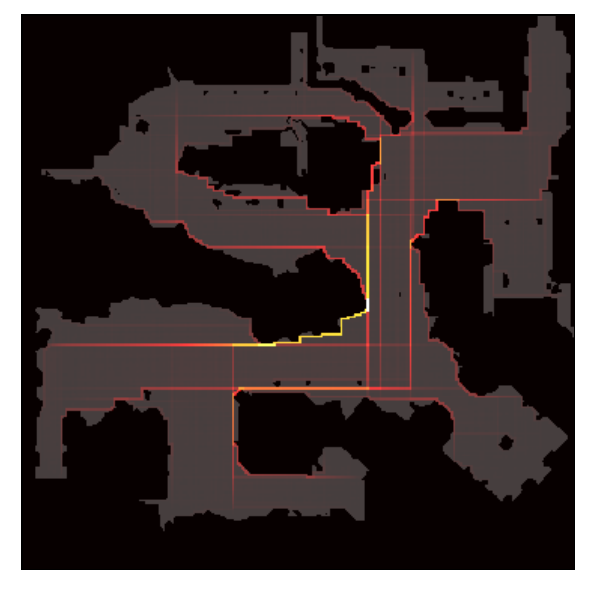}
    \caption{CBSH}
    \end{subfigure}
    \begin{subfigure}{.15\textheight}
    \centering
    \includegraphics[width=\columnwidth]{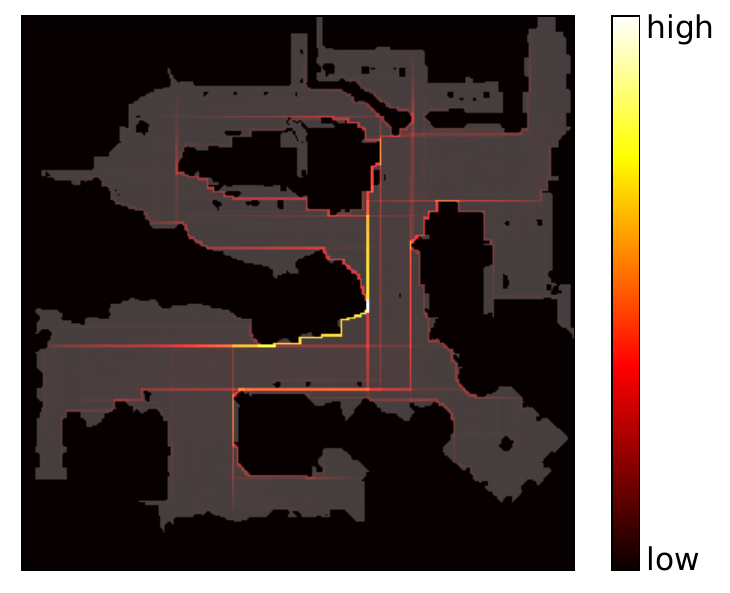}
    \caption{BCP}
    \label{fig:heatmap_6}
    \end{subfigure}
        \caption{(a)-(d) The scatter plots for average single-agent shortest path length and number of agents with respect to SpaceRatio. (e)-(j) Heat maps of the single-agent shortest paths with respect to different algorithms for (e) - (g) Berlin and (h) - (j) den520d.}
    \label{fig:analysis}
    \Description{Supplemental analysis of different MAPF algorithms on two difference map, Berlin and den520d. It shows that the number of agents and SpaceRatio of the instances that a specific algorithm ranks as fastest are different. The CBS algorithms tends to work better for shorter single-agent shortest paths and vice versa for CBSH and BCP.}
\end{figure*}

\subsection{Dataset Analysis} \label{sec:data_analysis}

In order to gain a deeper understanding of our MAPF algorithm portfolio, we analyze the performance of each algorithm for all the MAPF instances we generated for training and testing the algorithm selector.
Although interpreting why MAPFAST chooses a certain algorithm is beyond the scope of this paper, we aim to provide more insight on when a specific algorithm might work well for a certain scenario.

Since the input for MAPFAST contains the single-agent shortest paths, there may be some corresponding patterns of these paths for the test cases that have the same fastest algorithms. 
Here we define \emph{SpaceRatio}, which is equal to the number of map cells that are on the single-agent shortest paths divided by the number of the map cells that have no obstacles in it. 
The SpaceRatio not only represents how much free map space is used by the single-agent shortest paths, but also how spread the start and goal locations are in a map. 
We present the scatter plots for average length of single-agent shortest paths with respect to SpaceRatio of two different maps in Fig.~\ref{fig:apl_sp_berlin} and ~\ref{fig:apl_sp_den520d}. 
Each data point is colored by the algorithm that solves the corresponding instance fastest.
The distributions of the average single-agent shortest path lengths and SpaceRatio with respect to each algorithm are also shown on the top and right sides of the figures. 
We see that CBS tends to perform better than CBSH and BCP for the test cases having shorter average length of single-agent shortest path and smaller SpaceRatios. 
CBSH and BCP have similar performance on different average single-agent shortest path lengths. 
However, CBSH performs better than BCP on the test cases with higher SpaceRatios. 
Since the SpaceRatio is also affected by the total number of agents, we further present the scatter plots for the number of agents with respect to SpaceRatios in Fig.~\ref{fig:ag_sp_berlin} and Fig.~\ref{fig:ag_sp_den520d}. 
When there are fewer agents in the map, CBS works better for smaller SpaceRatios while BCP and CBSH dominate the test cases with larger SpaceRatios.

In Fig.~\ref{fig:heatmap_1} - \ref{fig:heatmap_6}, we show the heat maps of the single-agent shortest paths for all of the test cases where a certain algorithm is ranked as the fastest one. 
The more a map cell is used by a single-agent shortest path, the brighter it is. 
The heat maps for CBS (Fig.~\ref{fig:heatmap_1} and \ref{fig:heatmap_4}) have lots of scattered shortest paths compared to BCP and CBSH. 
This is because the solving speed of CBS is determined by the number of conflicts found during the search phase. 
The test cases with longer single-agent shortest paths tends to result in more potential conflicts, thus making CBS slower.
On the other hand, the heat maps for CBSH and BCP are mostly dominated by the longer paths. 
Although it seems that the heat maps for CBS have occupied more map space than CBSH and BCP, it has no correlation with the SpaceRatio since the heat maps contain paths from different test cases. 
The notable differences of CBS's heat maps with other algorithms also demonstrate our motivation of adding single-agent shortest paths as an input tensor for MAPFAST. 
The differences in the heat maps are so readily apparent that a human can manually decide whether or not to use CBS without the help of an algorithm selector. 
However,  the heat maps alone do not lead to any obvious suggestion about when to use BCP or CBSH. 
These two algorithms have very similar performance on test cases with different numbers of agents and SpaceRatios. 
Although the test cases in Fig.~\ref{fig:analysis} indicate that CBSH works better for higher SpaceRatios, we have observed similar results for BCP in other maps which are not shown here.

Based on the dataset analysis, one interesting future work is to develop \emph{hybrid} MAPF algorithms that combine the strength of different algorithms. 
For instance, one can use the number of agents or SpaceRatio as an additional heuristic to help decide whether to use the basic version of an algorithm such as CBS or an improved version such as CBSH. 


\section{Conclusions and Future Work}

In this paper, we present MAPFAST, a deep learning based optimal MAPF algorithm selector that outperforms the current state-of-the-art model.
We also introduce a new encoding method for the MAPF instances by using the single-agent shortest paths. 
In addition to just using classification loss in the CNN model, we show that adding multiple supplemental loss functions such as completion loss and pair-wise loss further improves the performance of the algorithm selector. 
The performance for MAPFAST is evaluated and analyzed with a large and diverse dataset of MAPF instances. We empirically show that the single-agent shortest paths, even without map topology, contain enough information to train a model that outperforms all portfolio algorithms as well as the state-of-the-art model. 
Also, we provide insight on the inherent features of MAPF problems that may help future researchers improve their MAPF algorithm designs.

We propose the following problems for future work: (1) utilizing graph-based learning techniques to extend to MAPF on general graphs, (2) incorporating our algorithm selection model into MAPF problem decomposition to select an efficient solver for each sub-problem, and (3) incorporating sub-optimal MAPF algorithms into our portfolio and training a model that selects fast solvers with near-optimal cost.



\begin{acks}
This research was supported by NSF awards IIS-1553726, IIS-1724392, IIS-1724399, and CNS-1837779 as well as a gift from Amazon.
\end{acks}



\bibliographystyle{ACM-Reference-Format} 
\bibliography{mybib}


\end{document}